\documentclass[12pt,preprint]{aastex}

\shorttitle{Helioseismology and solar abundances} 
\shortauthors{Antia and Basu}

\begin{document}

\title{Determining solar abundances using helioseismology}

\author{H. M. Antia}
\affil{Tata Institute of Fundamental Research, Homi Bhabha Road,
Mumbai 400005, India}
\email{antia@tifr.res.in}

\and

\author{Sarbani Basu}
\affil{Astronomy Department, Yale University, P. O. Box 208101,
New Haven CT 06520-8101, U.S.A.}
\email{sarbani.basu@yale.edu}

\begin{abstract}
The recent downward revision of solar photospheric
abundances of Oxygen and other heavy elements has resulted in serious
discrepancies between solar models and solar structure
as determined through helioseismology.
In this work we investigate the possibility of determining
the solar heavy-element abundance without reference to spectroscopy by using 
 helioseismic data.
Using the dimensionless sound-speed derivative
in the solar convection zone, we find that the heavy element
abundance, $Z$, of $0.0172\pm0.002$, which is closer to the older, higher value of 
the abundances.
\end{abstract}


\section{Introduction}

Recent analyses of spectroscopic data 
have suggested that the solar abundance of Oxygen and other abundant elements
needs to be revised downward (Allende Prieto, Lambert \& Asplund 2001, 2002).
 Asplund et al.~(2004, 2005) claim that the solar
oxygen abundance should be reduced by a factor of about 1.5  from the earlier estimates of
Grevesse \& Sauval (1998; henceforth GS). The abundances of C, N, Ne, Ar and other
elements are also reduced (Asplund et al.~2004, 2005; henceforth ASP), 
which causes the heavy element abundance
in the solar envelope to reduce from $Z=0.017$ to 0.0122. Solar models
constructed with this low heavy element abundance are found to have too shallow
a convection zone, and a helium abundance $Y$ that is too low. The
sound-speed and density profiles of the models also do not match the 
seismically inferred profiles
(Bahcall \& Pinsonneault 2004; Basu \& Antia 2004; Bahcall et al.~2005a, 2005c;
etc.).

Attempts to resolve the problem by increasing the rates of diffusion of helium and heavy elements
failed (Montalb\'an et al.~2004; Guzik, Watson \& Cox 2005; Basu \& Antia 2004)
since the resulting models had envelopes with much lower helium abundances than that observed.
Basu \& Antia (2004) and Bahcall et al.~(2005a)  found that
increasing  the opacities by 15--20\% near the convection zone base
can resolve the discrepancy between the models and the seismically inferred
profiles. However, a recalculation of
the opacities by the Opacity Project (OP) group (Seaton \& Badnell 2004; Badnell et al.~2005) showed no
significant increase in the
opacities near the base of the convection zone, and therefore the models remain discrepant.
In fact Bahcall, Serenelli \& Basu (2005d) find that even if all the input parameters  i.e.,  opacities, 
nuclear reaction rates,  equation of state (EOS), and diffusion rates are
changed within their estimated errors, 
solar models with the ASP abundances do not agree with the Sun. 
Similarly, Delahaye \& Pinsonneault (2005) show that the low abundances
strongly disagree with standard solar models at a level of $15\sigma$.

Antia \& Basu (2005) and Bahcall, Basu \& Serenelli (2005b) proposed
 that an increased
Neon abundance 
could resolve the discrepancy. Measurements of Ne abundance
are based on coronal lines and may not, as in the case of helium, represent photospheric
abundances.
This solution seemed to be
particularly promising since Drake \& Testa (2005) found that most neighboring stars seem to
have a much higher Ne/O ratio as compared to the Sun.
However, Schmelz et al.~(2005) and Young (2005) 
reanalyzed solar X-ray and UV data respectively to find that the Ne/O ratio of the Sun is indeed
consistent with the old lower value.
Interestingly, recent measurement of Oxygen abundance in the solar
atmosphere using CO lines by Ayres et al.~(2005) using the Shuttle-borne
Atmospheric Trace Molecule Spectroscopy (ATMOS) Fourier transform spectrometer combined with some ground based
observations, yields an Oxygen abundance that is much higher than the recent ASP
value and close to that of GS. If this result is confirmed, then the
discrepancy between the solar models and seismic constraints will be
automatically resolved.

All the seismic constraints so far have been obtained by comparing solar models
with seismically inferred convection zone depth and envelope helium abundance.
In these models the effect of the heavy element abundance manifests mainly through the
opacity which controls the structure of radiative interior. Since the
structure of the radiative interior also depends on other input physics, like
nuclear energy generation rates, diffusion coefficients, etc., it is
difficult to isolate the effect of heavy element abundances on these models.
Although Bahcall et al.~(2005d) and
Delahaye \& Pinsonneault (2005) have made detailed estimates of possible
uncertainties in input physics, one cannot rule out the possibility that
some effect has been inadvertently ignored or underestimated. Hence it would be interesting
to get an independent estimate of the solar heavy element abundance from seismic
data using only the structure of the solar convection zone. The structure of the
solar convection zone  is independent
of many of the above mentioned uncertainties in input physics, and depends mainly
on the equation of state. In fact,
Basu \& Antia (2004) and Antia \& Basu (2005) used solar envelope models
to circumvent these uncertainties. However, the convection-zone structure
in these models also depends somewhat on the opacities near the base of the
convection zone, and hence,
these envelope models
are also not completely free from these uncertainties.

In this work we investigate the possibility of determining the heavy-element
abundance of the Sun using helioseismic data in the manner in which solar
helium abundance has been determined (e.g., D\"appen et al.~1991; Antia \& Basu 1994). The process
of ionization lowers the  adiabatic index
$\Gamma_1$ in the ionization zones thereby lowering the sound speed. However,
the sound speed increases too rapidly with depth because of the increase in temperature to display this modulation easily. Gough (1984)
pointed out that we can use the sound-speed gradient to study the modulation of the adiabatic 
index. Gough (1984) defined a dimensionless sound-speed gradient 
\begin{equation}
W(r) ={r^2\over Gm} {dc^2\over dr},
\label{eq:wr}
\end{equation}
where $m$ is the mass inside the
spherical shell of radius $r$, $G$ is the gravitational constant and $c$ the speed of sound.
In an adiabatically stratified region, such as most of the solar convection zone, $W(r)$ is related to
adiabatic indices.
If gas were fully ionized below the helium ionization zone, 
$W(r)$ would be $-2/3$. However, this is the region where many of the heavy elements are ionizing,
and hence $W(r)$ deviates from $-2/3$ and this deviation can be used to
measure the heavy element abundances.
Although different elements leave separate imprints on $W(r)$,
these are small (because of low abundances), and hence unlike in the case of helium, it will be difficult to
use $W(r)$ to determine the abundance of each element individually. 
In this work we investigate the possibility of estimating the total solar heavy element
abundance using $W(r)$. This technique requires the inversion of solar
oscillation frequencies to
determine solar  structure
in the lower part of the convection zone, and the inversions are known to be  very reliable. Furthermore,
the signature of heavy element abundance depends essentially on the equation
of state and is essentially independent of other input physics in the solar model.

The rest of the paper is organized as follows:
In Section~2, we identify the signature of heavy element abundance on the
profile of $W(r)$, while in Section ~3, we show how this signature can be
utilized by structure inversion to estimate $Z$. We discuss the significance
of these results in Section~4.

\section{The signature of heavy elements}
\label{sec:wr}

\begin{figure}
\plotone{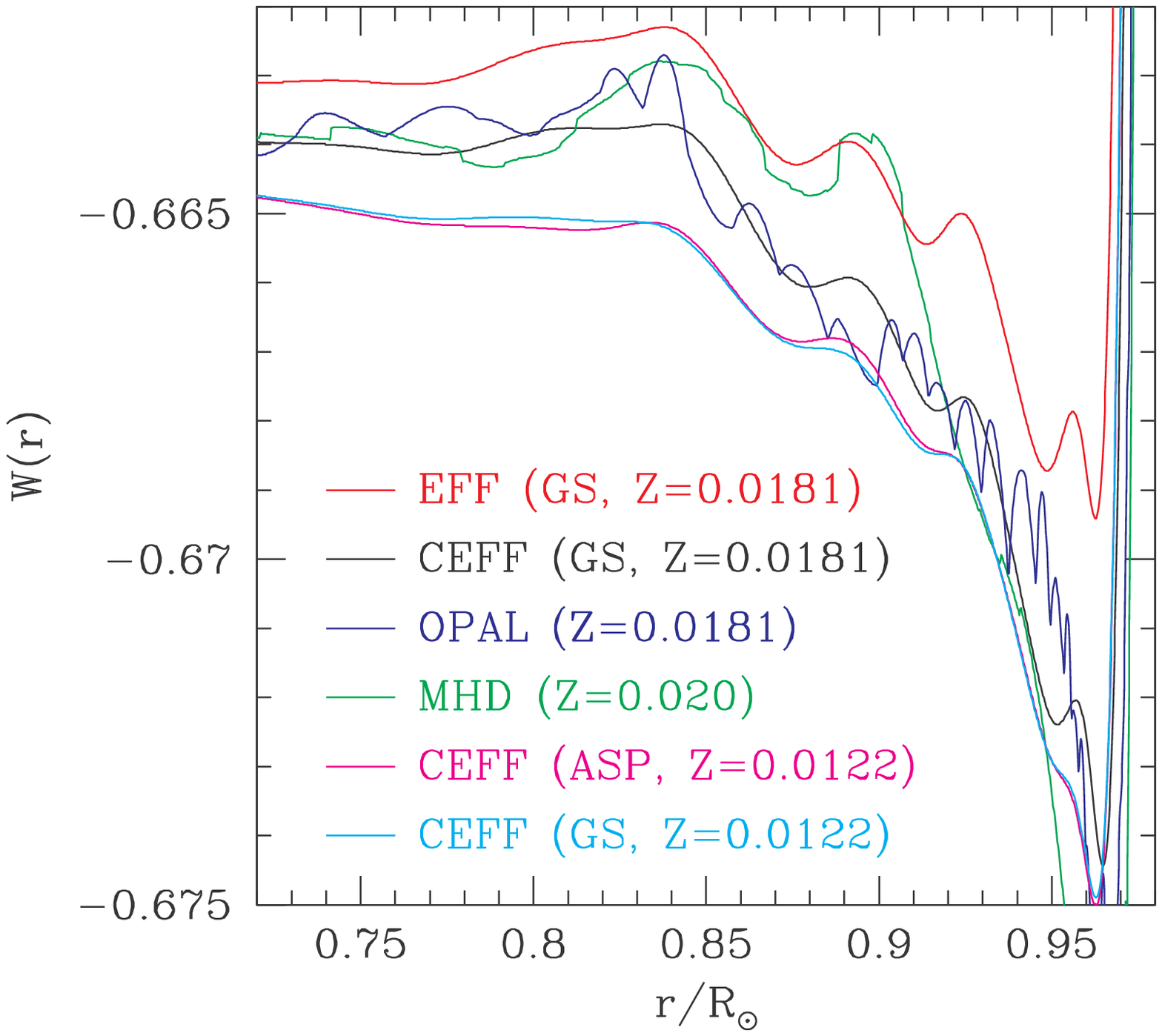}
\caption{The function $W(r)$ for solar envelope models
constructed with different equations of state, heavy-element mixtures, and abundances as marked in the
figure. All these models have their convection zone base at $r_b=0.7133R_\odot$,
and have a hydrogen abundance $X=0.739$.
\label{fig:wrall}}
\end{figure}

In order to study the influence of heavy elements on $W(r)$ we construct
solar envelope models with different abundances. Since the equation of state tables of
OPAL (the Opacity Project At Livermore; Rogers \& Nayfonov 2002) and MHD (``Mihalas, Hummer \& D\"appen'';
D\"appen et al.~1987, 1988; Hummer \& Mihalas
1988; Mihalas et al.~1988) are available only for a fixed mixture of heavy elements, we
use the Eggleton, Faulkner \& Flannery (1973, henceforth EFF) equation of state with coulomb
corrections (Guenther et al.~1992; Christensen-Dalsgaard \& D\"appen 1992).
This equation of state (referred to as CEFF, the `Coulomb-corrected' EFF) includes all ionization states 
of 20 elements.
The ionization fractions are calculated using only the ground state partition function.
Fig.~\ref{fig:wrall} shows $W(r)$ for solar models with different $Z$ and
different mixtures (GS or ASP). For comparison, models using OPAL, MHD and
EFF (i.e., without applying the coulomb corrections) are also shown.
Fig.~\ref{fig:wrall} shows clearly the
equation of state dependence of $W(r)$ for models with the same $Z$.
The function $W(r)$ of the solar models with OPAL equation of state shows some
oscillations. We believe that these are caused by interpolation in tables
that have limited precision --- quantities like $\Gamma_1$ etc.,
are listed only to four decimal places in the OPAL tables.
It is known that the OPAL equation of state is very close to the equation of state in the Sun (e.g., Basu \& Antia 1995;
Elliott 1996; Basu \& Christensen-Dalsgaard, 1997), and since from  Fig.~\ref{fig:wrall} we see that CEFF models have
$W(r)$ similar to that of OPAL models, we conclude that CEFF models can also be used for calibration. 
The effect of equation of state on the inferred $Z$ can be estimated using test models (see \S~\ref{sec:sun} for details).

\begin{figure}
\plotone{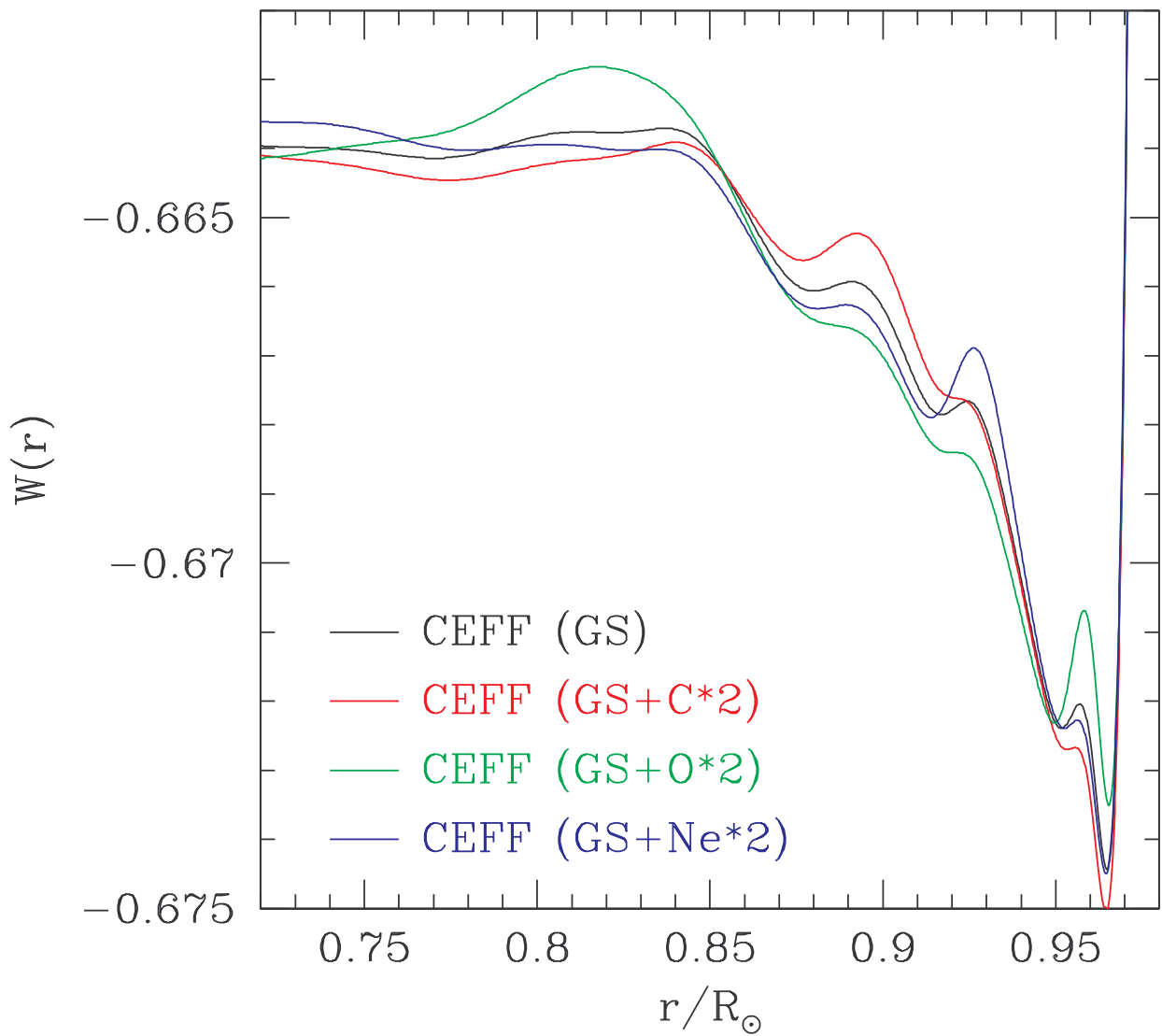}
\caption{The function $W(r)$ for solar envelope models with heavy element abundance $Z=0.0181$ constructed 
with CEFF equation of state but for different heavy-element mixtures as marked in the
figure. These mixtures include cases where the relative abundance of C, N or O are
increased by a factor of 2 over the GS value.
All these models have the convection zone base at $r_b=0.7133R_\odot$,
and have a hydrogen abundance $X=0.739$.
\label{fig:wrz}}
\end{figure}

We find that the differences in $W(r)$ due to differences in $Z$ are very clear. The
differences in $W(r)$ for models with the same $Z$ but different relative abundances of
heavy elements, i.e., different mixtures, is more subtle and less obvious. 
We find that the models using GS and ASP mixtures normalized to the same value
of $Z$ are close to each other.
This could be
because ASP have revised the abundance of the most abundant
elements by the same factor. 
Thus while we
may not be able to determine the mixture of elements in the Sun, but we can
hope to determine $Z$, the total heavy element abundances. 
Nevertheless, it may be possible to determine abundances of abundant elements
like Carbon, Oxygen and Neon if reliable equation of state is available. Figure~\ref{fig:wrz} shows
$W(r)$ for a few models constructed using different mixtures, and includes
some models where the relative abundances of C, O or Ne have been increased by a
factor of two as compared to the GS value. In these models,  the abundance of
one element is increased by a factor of two relative to the GS mixture, and then the resulting
mixture is scaled to the same value of $Z$.
All these models are based on the CEFF
equation of state since the  other, more sophisticated, equations of state do not allow different mixtures to be used.
We can identify some peaks in $W(r)$ caused by  C, O, and Ne  around
$r=0.89R_\odot, 0.82R_\odot$ and $0.92R_\odot$ respectively, in these curves.
For different  equations of state, the ionization zones
may shift to some extent and hence these peaks can be used to determine
abundances only if a reliable equation of state is available to pin down
the positions of these peaks.

\begin{figure}
\epsscale{.80}
\plotone{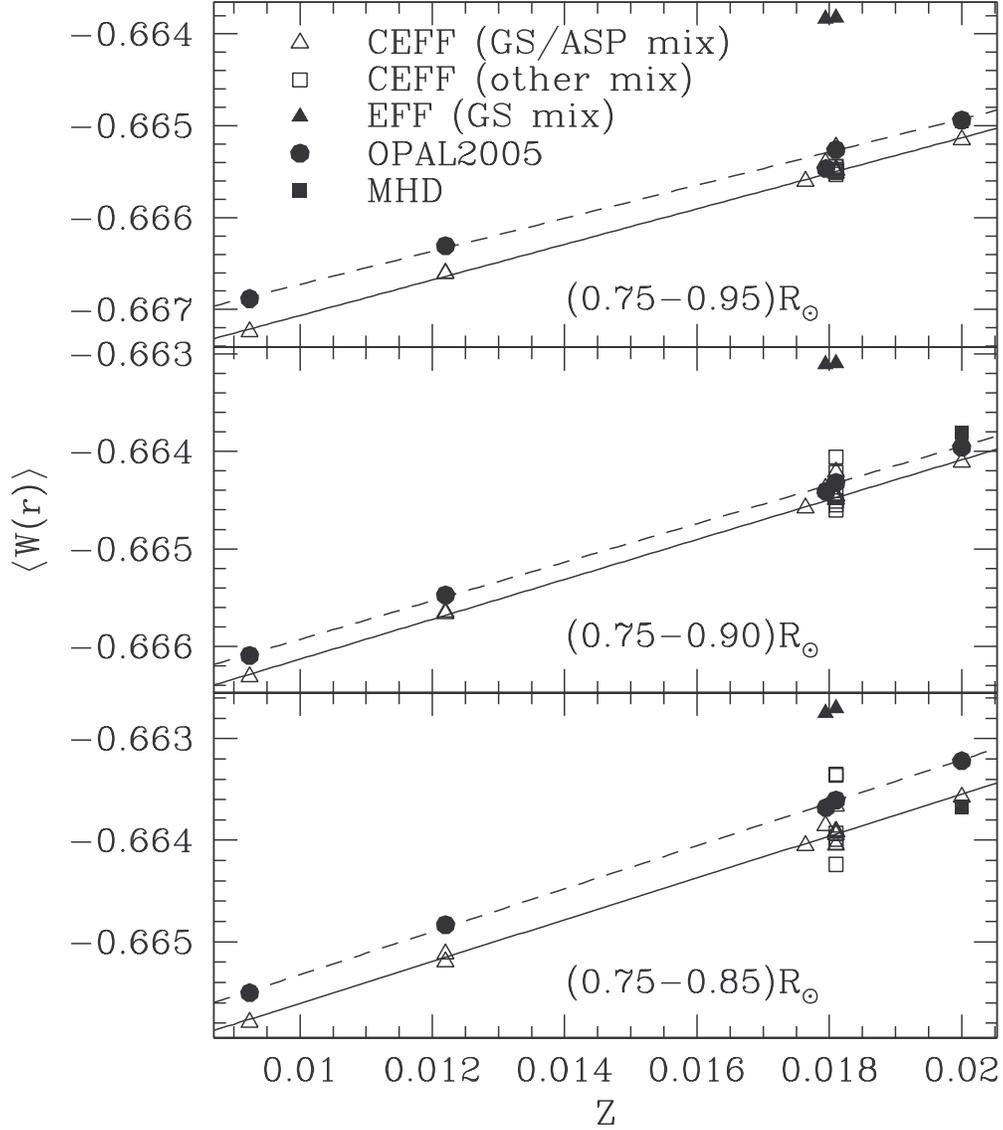}
\caption{The average value of $W(r)$ in three different radius ranges 
plotted as a function of $Z$ for solar models with different equations of state and mixtures
of heavy elements as marked in the figure.
The CEFF models with `other mixtures' include models constructed with the
abundance of some individual elements increased by a factor of 2 as compared
to GS value. These models are listed in Table~1.
The reference and test models used in this study are also included in this figure.
The continuous line is the best-flt line through CEFF models of different $Z$,
the dashed line is that through OPAL models, and these lines are used to determine
$Z$ of the test models and the Sun using inverted $\langle W(r)\rangle$.
\label{fig:wint}}
\end{figure}

We quantify  $W(r)$ by calculating the average value
of $W(r)$ in different radius ranges
\begin{equation}
\langle W(r)\rangle= {\int_{r_1}^{r_2} W(r) dr\over r_2-r_1}\;,
\label{eq:int}
\end{equation}
and plot it in
Fig.~\ref{fig:wint}. It is clear that $\langle W(r)\rangle $ in all three radius
ranges increases almost linearly with $Z$. The best fit straight lines shown
in the figure are those obtained using models with CEFF (with GS mixture) and OPAL equation of state, and
shows that by calibrating $\langle W(r)\rangle $ using the fitted relation,
we can determine $Z$. The various models around $Z=0.0181$ in Fig.~\ref{fig:wint}
correspond to different mixtures and other parameters of the  solar models.
Since it is not possible to distinguish between different symbols in the
figure, the corresponding differences in $\langle W(r)\rangle$ are
listed in Table~1.
The reason for 
averaging $W(r)$ over different ranges in radius is to check the sensitivity of the estimated $Z$ to differences
arising due to the locations of the ionization zones of different elements. Furthermore,
it turns out that inversions are not very reliable in the outer regions of the Sun,
and hence it may be necessary to avoid that region in some cases.
However, looking at the distribution of points for the different mixtures in
Fig.~\ref{fig:wint} it appears that over the range $(0.75,0.95)R_\odot$
the results appear to be less sensitive to differences in mixture and are
determined by essentially the total heavy element abundance. Thus while in
the inner radius range the inversions are more reliable, the results in this range are
more sensitive to differences in mixtures. This may not be important for
distinguishing between GS and ASP mixtures since in both these mixtures the relative
abundances of C, N, O, Ne (which are most relevant in the convection zone)
are similar.  In principle,
by considering $W(r)$ in different radius ranges it should be possible to
determine abundances of individual elements, but that would require better
inversions in the outer regions and also a more reliable equation of state that can be calculated with different heavy element mixtures.
The maximum effect is obtained by changing the relative abundance of O
and the effect is maximum in the deeper radius ranges. Thus if the O abundance is
increased, the value of $\langle W(r)\rangle$ increases in all radius
ranges, which would imply that the estimated value of $Z$ using the
calibration curves for standard mixture will tend to overestimate $Z$.
Furthermore, the difference will  be the most in the deeper radius range.
The abundance of C has a
smaller effect, and   is opposite in sign as compared to O,
presumably because C gets completely ionized in the convection zone.
It may be noted that a change by a factor of 2 in O abundance is much
more than the difference between GS and ASP abundances and the only reason
for choosing this variation was to accentuate the possible signatures of the
abundances of individual elements. Thus the variation between different
models in Fig.~3 does not represent typical errors that may be expected
from uncertainties in the mixture of heavy elements. Furthermore, if the C abundance
is also increased along with the abundance of O, then some of the difference will get canceled.

Other potential sources of systematic errors are differences in the
hydrogen abundance, $X$, the depth of the convection zone, 
atmospheric opacities,  treatment of convection, and  solar radius.
The effect on solar envelope models of differences in opacities near the convection zone base is the same
as that of changing the convection zone-depth.
We have checked
for sensitivities to each of these factors by constructing solar models with different
inputs and the results are shown in Table~1. All differences in this table
are taken for a fixed value of $Z=0.0181$ with respect to a ``standard model''
using CEFF equation of state with GS mixture. To check for the influence of atmospheric
opacity we use OPAL (Iglesias \& Rogers 1996) instead of Kurucz opacities (Kurucz 1991) at low temperatures. 
Similarly, to test for effect of treatment of convection we use the mixing
length theory instead of formulation due to Canuto \& Mazzitelli (1991).
The largest difference (apart from that due to equation of state and different mixtures) is due to the convection zone 
depth --- a difference of $0.01R_\odot$ in the depth of the convection zone
changes $\langle W(r) \rangle $ by about $2.5\times10^{-4}$. Although,
the depth of the convection zone is known to an accuracy of $0.001R_\odot$
(Basu \& Antia 1997, 2004) we have considered much larger perturbations to
account for the effects of varying opacities near the base of the convection
zone. The opacity variations may arise due to uncertainties in $Z$.
The shift in the convection-zone base of $0.01R_\odot$ is a little more than half the shift caused by
reducing $Z$ from the GS value  to the ASP value.

\section{Determining $Z$ in the Solar Convection Zone}
\label{sec:sun}

We have used solar oscillations frequencies obtained by the Global 
Oscillation Network Group (GONG) (Hill et al.~1996) as well as those from
the Michelson Doppler Imager (MDI) (Schou 1999) on board the Solar and Heliospheric Observatory (SOHO).
We use 97 sets of frequencies from GONG spanning the period from 1995 May 7 to 2005 Jan 1.
Each GONG frequency set was obtained from 108 days of observation.
We use 42 sets of frequencies from MDI covering the period 1996 May 1 to 2005 Jan 1.
Each MDI set was obtained by analyzing 72 days of observations.

\begin{figure}
\plotone{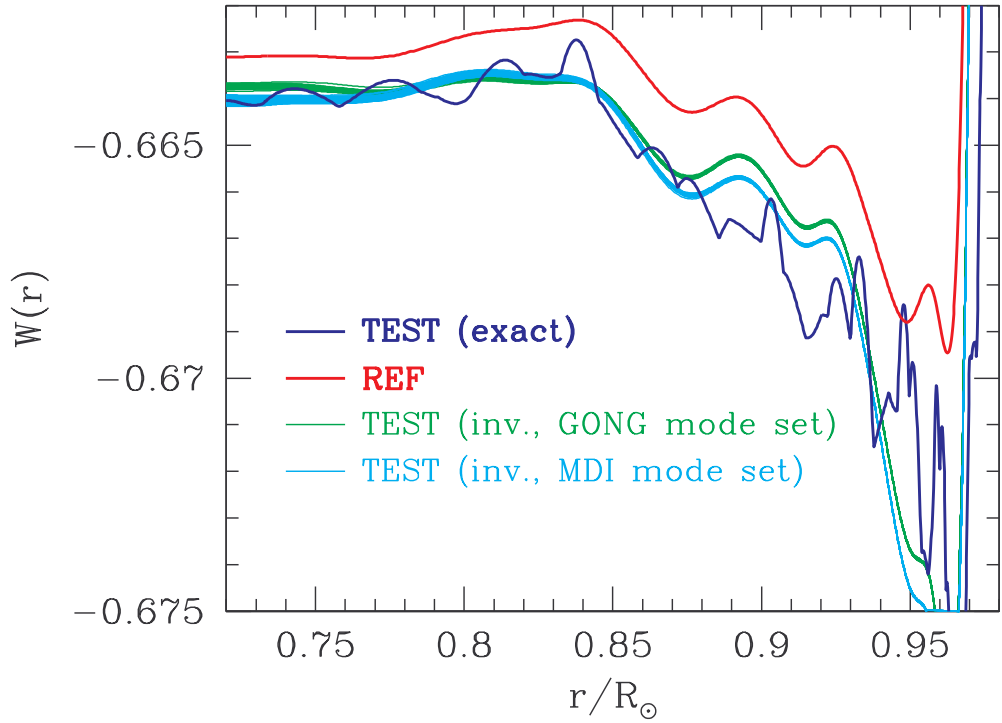}
\caption{The actual $W(r)$ for a test model constructed with the OPAL equation of state  compared
to  that inferred from inversions using an EFF model as the reference model.
The inversion results using 50 realizations of random errors from GONG and MDI
mode-sets are shown. The blue/green line shows the result for each of the
50 realizations for the GONG and MDI sets. Since the lines are close together,
the result appears as one thick line.
\label{fig:test}}
\end{figure}

We use the Regularized Least Squares (RLS) method (Antia 1996) to invert
the frequency differences between the Sun and a reference model. This method is
used since the solar
sound speed profile is obtained in a form that can be easily differentiated to calculate $W(r)$.
To test the reliability of helioseismic inversions for $W(r)$, we construct
two test solar models using the OPAL equation of state, and attempt to infer their $W(r)$ using a
reference model constructed with the EFF equation of state, which has a significantly
different  $W(r)$. One of these test models is a standard solar model with the
GS mixture of heavy elements,  the other is an envelope model with  a lower $Z$, but the same
convection zone depth as the standard model. 
In order to simulate the effect of observational errors, we have
used only those modes that are present in the observed mode-sets.
We use one mode-set each
from GONG and MDI datasets, and for each set we obtained 50 different
realizations of random errors that were added to the frequencies of the
models. The results obtained using all these realizations for the first test
model
are shown in Fig.~\ref{fig:test}. Each line in this figure shows the
results for one realization of errors. It is clear that we can successfully invert 
for $W(r)$ although the error in inferred $W(r)$ tends to increase with increasing $r$, an
expected result since the mode-sets used do not have many modes that are sensitive to solar
structure at larger $r$.
The inversions are more reliable in the lower half of the convection zone.
Furthermore, using $\langle W(r)\rangle$
for different radius ranges we can estimate the value of $Z$ using 
the  best fit straight lines in Fig.~\ref{fig:wint}. The results are listed in Table~2.
The errors listed in the table are caused by errors in the observed frequencies, and are just
the standard deviations of the results estimated from the 50 sets of error-realizations
for each data set and test model.
The inferred value of $Z$ is slightly larger than the actual value in the
model when $\langle W(r)\rangle$ is calibrated using CEFF models. 
This is a result of the equation of state related difference in  $W(r)$:
OPAL equation of state tends to give larger $W(r)$ for the same $Z$ than
models with CEFF equation of state (see Fig.~\ref{fig:wint}).
If OPAL models are used for calibration, then we get a result that
is close to the actual value.
A systematic error of the order of 0.0015 may be expected due to differences
in equation of state. The mean value of $Z$ using CEFF calibration models
is $0.0191\pm0.0004$, while that using OPAL calibration models is
$0.0175\pm0.0005$, where in both cases the errors are the standard deviation
of the 6 values in Table~2. If all 12 results in Table~2 are averaged we
get $Z=0.0183\pm0.0009$.
Some of these differences between results using OPAL and CEFF calibration
models are due to  differences in the heavy element mixture
used in CEFF and OPAL equation of state calculations (see Table 1). If we construct CEFF models using
the same mixture as that in OPAL equation of state, the difference is somewhat reduced, but
we prefer to show the results using CEFF equation of state with GS mixture as that would
in some sense give the combined error that arises due to difference in both the equation of state and the heavy-element mixtures.
By considering inversion results with different test models,
we find that inversions with the MDI data set are better (i.e., they are
closer to the true values in the outer region) than those with the
GONG data set, particularly in region close to $r=0.95R_\odot$.
This problem is particularly serious for the test model with low $Z$, which
has a significantly different density profile as compared to reference model.
This is most likely due to the fact that GONG sets do not have any modes
with degree larger than 150 that are required to get good inversions above 
$0.9R_\odot$. As a result, we have not shown the corresponding results in
Table~2.

We have done similar tests with  different test models and we find that a
reliable inversions of $W(r)$ is possible only if the test and reference
models have similar convection zone depths.
This is expected since 
$W(r)$ increases very steeply just below the convection zone base.
If the convection zone depth of the two models do not
match, the finite resolution of the inversions causes some leakage of the   reference model $W(r)$ into
the inverted $W(r)$  of the test model.
Since the change in $W(r)$ at the convection zone base is almost three orders of magnitude
larger than the vertical scale used in Fig.~\ref{fig:wrall}, it is impossible
to avoid an error  unless the depths of the convection zone match.
To avoid this error while determining  $Z$ inside the Sun, 
we use reference models that have nearly correct convection zone depths.

\begin{figure}
\plotone{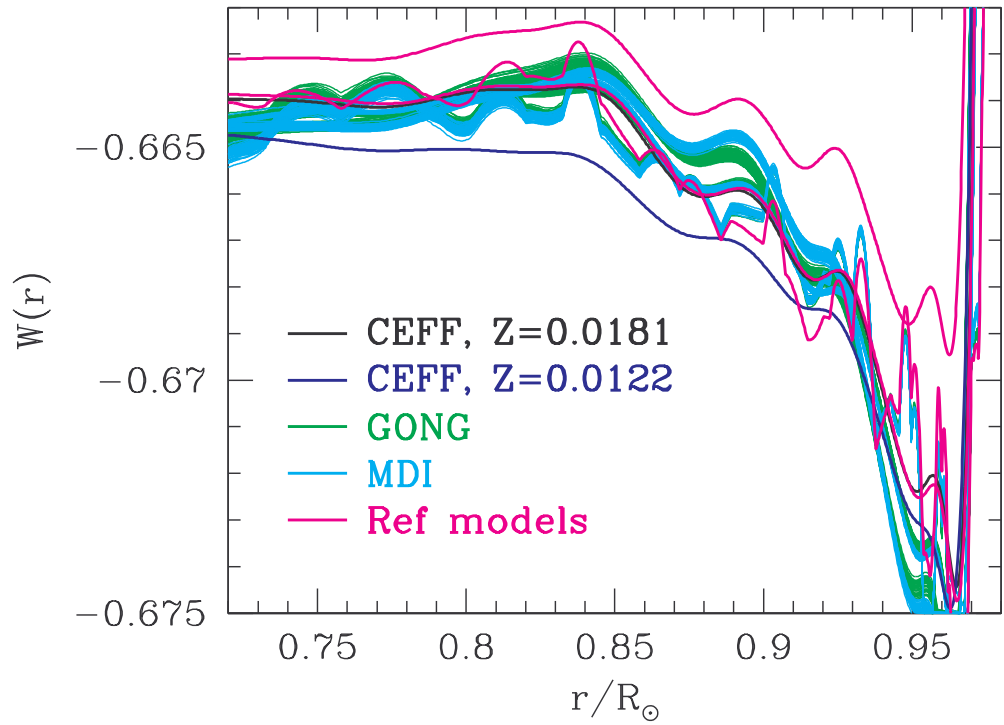}
\caption{The function $W(r)$ for the Sun obtained using three different reference
models for the different sets of solar oscillations frequencies from GONG and MDI are
compared with that in a solar model with different heavy element abundances.
\label{fig:obs}}
\end{figure}

The three reference models used to invert the solar frequencies have been constructed using EFF, CEFF and OPAL equations of state. To 
estimate the systematic error caused by differences in
structure between the reference models and the Sun, we use all these models as reference
models to invert each set of observed frequencies and calculate
$W(r)$ and the results are
shown in Fig.~\ref{fig:obs}. Also shown are $W(r)$ for models
with $Z=0.0181$ (approximately the GS value) and $Z=0.0122$ (the ASP value).
It is clear from the figure that the solar $W(r)$ supports the higher value of $Z$,
and that all the data sets and reference models give very similar results. 
Table~2 lists the value of $Z$ in the Sun inferred 
using CEFF and OPAL models for calibration.  The average $Z$ obtained using
the three different reference models is listed. The reference and calibration
models should not be confused --- the calibration models are used to
determine $Z$ for a given $\langle W(r) \rangle$, the reference models
are needed to invert solar oscillations frequencies to obtain $W(r)$ inside the Sun.
The errors quoted
in the table are the standard deviation of all inversion results obtained
using the different sets of solar oscillations frequencies, each set  inverted using the three different reference
models. 
Because of the use of multiple reference models, the estimated
error for the solar results is larger than that for test models.
Table~2 shows that all 12 values of $Z$ inferred using  observed solar
frequencies are close to GS value and much larger than the revised
estimate of ASP.

\section{Conclusions}
\label{sec:conclu}

We have examined the possibility of determining the solar heavy element abundance using
$W(r)$, the dimensionless gradient of sound speed,  obtained 
by inverting solar oscillations frequencies.
The value of this function below the HeII ionization zone is determined
by the heavy element abundance. 
Using test models we find that we can determine Z 
by comparing the  average value of $W(r)$ in the lower
convection zone to that of models with known $Z$. The mean value of $Z$ obtained by
averaging all values listed in Table~2 for the two test models are
$Z=0.0183\pm0.0009$ and $Z=0.0128\pm0.0009$, respectively.

For each  set of observed frequencies obtained from the GONG and MDI projects, we calculate
$W(r)$ with three different reference models. The average value of
$W(r)$ in three different radius ranges is used to estimate the value of $Z$
using either CEFF or OPAL calibration models. The mean of all these
measurements yields the solar solar heavy element abundance 
of  $Z=0.0172\pm0.002$,
which is similar to the value obtained by GS
and disagrees with the reduced abundances as determined by ASP at about the $2.5\sigma$ level.
The errorbar accounts for the systematic
errors caused by differences in the equation of state and other parameters of the calibration and reference
solar models used.
The standard deviation of the 12 values listed in Table~2 is 0.0009, which
should include the differences due to equation of state as two different sets of calibration
models using CEFF and OPAL equation of state are used for estimating $Z$ from $\langle W(r)\rangle$.
Similarly, possible systematic errors due to differences in mixture should also
contribute to this standard deviation, since in that case $\langle W(r)\rangle$
for the different depth ranges will give different results.
The value of $\langle W(r)\rangle$ in the lower convection zone is not particularly sensitive
to the equation of state used, and models constructed with equations of state other than the  EFF equation of state (which is known to be deficient)
give similar results.
Using test models we find that the equation of state (and $Z$ mixture) differences cause an error of 
0.0015 in $Z$ (difference between OPAL and CEFF values),
and a difference of $0.01R_\odot$ in the convection zone depth 
translates to an error of 0.0012 in $Z$, which gives the quoted error of 0.002. 
These error estimates are probably conservative since, for example, the difference
of $0.01R_\odot$ in the convection zone depth is much larger than any realistic expectation even
if we include the effect of uncertainties in opacities. Similarly, the error in $Z$
due to the equation of state should be normally taken as half the difference caused by two
different equations of state (e.g., Delahaye \& Pinsonneault 2005). The error due to
differences in mixture, if considered separately, would be about 0.001.
However, since it is difficult to
estimate systematic errors, we have adopted an estimated error of 0.002 for
 this work. The error estimate is comparable to that in spectroscopic
determination of abundances.

Similar estimates for $Z$ have been previously obtained by using the current OPAL opacities.
Basu \& Antia (1997), Basu (1998), Antia \& Basu  (2005)   found that to obtain solar models with
the correct convection zone depth and density profile the required $Z$ needs
to be close to 0.017 if the OPAL opacities are valid near
the convection zone base. Delahaye \& Pinsonneault (2005) also found similar values
from a  detailed analysis of standard solar models.
Elliott (1996)  found $Z=0.0165$ using the
OPAL equation of state by inverting for the adiabatic index in the convection zone. 
Incidentally, he also
states that the Ne abundance needs to be increased beyond the then accepted value.
Thus using different and independent techniques
we get similar estimates for $Z$. 
If the
ASP abundances are correct, then not only the opacities but
the equation of state will also need substantial revision. It may be noted that all the currently
used equations of state and opacities (OPAL and OP) give similar results
for the solar heavy element abundance, and the results conflict with the new spectroscopically
estimated values. The results in this work are not very
sensitive to equation of state as a relatively crude equation of state like CEFF gives results comparable
to that using more sophisticated OPAL equation of state.

The drawback of the method used in this paper is that it is not sensitive to the relative abundances of
individual heavy elements in the GS or ASP mixtures. Hence it is not possible for us to identify
which element abundance needs to be increased as compared to the 
estimate of ASP.
The detailed profile of $W(r)$
depends on the relative abundances of heavy elements. Using models we infer that in
the convection zone below the He ionization zones $W(r)$ 
is essentially determined by abundances of C, O and Ne.
In order to determine the abundances of individual heavy elements,
we will need more sophisticated  equation of state tables
with higher accuracy for different heavy element mixtures.
We also need a better understanding of systematic errors in this
technique to enable us to determine the abundances of individual
elements.

\acknowledgments
This work utilizes data from the Solar Oscillations
Investigation / Michelson Doppler Imager (SOI/MDI) on the Solar and
Heliospheric Observatory (SOHO). The MDI project is supported by NASA
grant NAG5-8878 to Stanford University. SOHO is a project of international
cooperation between ESA and NASA. This work also  utilizes data obtained by the Global Oscillation Network
Group (GONG) project, managed by the National Solar Observatory, which is
operated by AURA, Inc. under a cooperative agreement with the NSF.
The data were acquired by instruments
operated by the Big Bear Solar Observatory, High Altitude Observatory,
Learmonth Solar Observatory, Udaipur Solar Observatory, Instituto de
Astrof\'{\i}sica  de Canarias, and Cerro Tololo Interamerican Observatory.
This work was partially supported by 
NSF grant ATM 0348837 to SB

\clearpage

\clearpage

\clearpage

\clearpage

\clearpage

\clearpage

\begin{deluxetable}{lrrr}
\tablecaption{Differences in $\langle W(r)\rangle$ due to various
sources of systematic errors in the sense (test model) $-$ (standard model
with CEFF equation of state). All models have $Z=0.0181$}
\tablehead{
\colhead{Model difference} & \multicolumn{3}{c}{$10^4\delta \langle W(r)\rangle$}\\
&\colhead{$(0.75,0.85)R_\odot$}& \colhead{$(0.75,0.90)R_\odot$}
&\colhead{$(0.75,0.95)R_\odot$}
}
\startdata
OPAL EOS& $3.859$& $1.274$& $2.076$\\
CEFF EOS with OPAL mixture& $3.187$& $0.101$& $-0.672$\\
CEFF EOS with ASP mixture& $-1.318$& $-0.342$& $-0.013$\\
Mixture with C increased by factor 2& $-3.218$& $-0.859$& $-0.228$\\
Mixture with N increased by factor 2& $-0.158$& $0.319$& $0.085$\\
Mixture with O increased by factor 2& $5.576$& $2.434$& $0.351$\\
Mixture with Ne increased by factor 2& $-0.734$& $-1.398$& $-0.455$\\
Mixture with Fe increased by factor 2& $-1.067$& $-0.476$& $-0.058$\\
Convection zone depth reduced by $0.01R_\odot$& $2.522$& $2.481$& $2.517$\\
With X reduced by 0.019& $-0.291$& $-0.270$& $-0.310$\\
With different low temperature opacity& $0.050$& $-0.024$& $-0.057$\\
With different treatment of convection& $-0.052$& $0.027$& $0.030$\\
With radius increased by 200 km& $-0.024$& $0.044$& $0.030$\\
      \enddata
    \label{tab:table1}
\end{deluxetable}

\clearpage

\begin{deluxetable}{lccccccc}
\tabletypesize{\scriptsize}
    \tablecaption{The heavy element abundance, $Z$ as inferred using
average values of $W(r)$ in different radius ranges for two test
models and for the Sun. The errorbars quoted in the table are the
respective standard deviation of all estimates and do not include other
systematic errors. The mean value for each test model and observed frequencies
is an average over all estimates listed here and the error bar is the standard
deviation of these values.}
\tablehead{
\colhead{Test} &  \colhead{Calib.} & \multicolumn{3}{c}{GONG data sets}&
\multicolumn{3}{c}{MDI data sets}\\
\colhead{model} & \colhead{model} & \colhead{$(0.75,0.85)R_\odot$} & \colhead{$(0.75,0.90)R_\odot$} & \colhead{$(0.75,0.95)R_\odot$} 
& \colhead{$(0.75,0.85)R_\odot$} & \colhead{$(0.75,0.90)R_\odot$} & \colhead{$(0.75,0.95)R_\odot$}
}
      \startdata
OPAL & CEFF & $0.0191\pm.0001$ & $0.0193\pm.0001$ & $0.0195\pm.0001$ &
 $0.0191\pm.0001$ & $0.0188\pm.0001$ & $0.0185\pm.0001$ \\
$Z=0.01795$ & OPAL & $0.0172\pm.0001$ & $0.0177\pm.0001$ & $0.0184\pm.0001$ &
 $0.0172\pm.0001$ & $0.0172\pm.0001$ & $0.0173\pm.0001$ \\
& & \multicolumn{6}{c}{\normalsize Mean value $Z=0.0183\pm0.0009$}\\
      \tableline
OPAL & CEFF & $0.0136\pm.0002$ & $0.0136\pm.0001$ & \nodata &
 $0.0136\pm.0002$ & $0.0128\pm.0002$ & $0.0140\pm.0001$ \\
$Z=0.0122$ & OPAL & $0.0120\pm.0002$ & $0.0123\pm.0001$ & \nodata &
 $0.0120\pm.0002$ & $0.0116\pm.0002$ & $0.0125\pm.0001$ \\
& & \multicolumn{6}{c}{\normalsize Mean value $Z=0.0128\pm0.0009$}\\
      \tableline
Obs. &  CEFF & $0.0179\pm.0004$ & $0.0183\pm.0009$ & $0.0181\pm.0003$ 
  & $0.0175\pm.0005$ & $0.0181\pm.0010$ & $0.0180\pm.0004$ \\
 &  OPAL & $0.0161\pm.0004$ & $0.0167\pm.0008$ & $0.0169\pm.0003$ 
 & $0.0157\pm.0005$ & $0.0166\pm.0009$ & $0.0168\pm.0004$ \\
& & \multicolumn{6}{c}{\normalsize Mean value $Z=0.0172\pm0.0009$}\\
\noalign{\smallskip}
      \enddata
    \label{tab:table}
\end{deluxetable}

\end{document}